\documentclass[osa,lengthcheck,superscriptaddress]{revtex4-1}

\usepackage{amsmath}
\usepackage{amssymb}

\usepackage{graphicx}

\begin{document}

\title{Enhancement of radiative processes in nanofibers with embedded plasmonic nanoparticles}

\author{Radoslaw Jurga}
\affiliation{Istituto Italiano di Tecnologia (IIT), Center for Biomolecular Nanotechnologies, Via Barsanti, 73010 Arnesano, Italy.}
\affiliation{Department of Mathematics and Physics ``Ennio De Giorgi'', University of Salento, Via per Arnesano, 73100, Italy.}

\author{Fabio Della Sala}
\affiliation{Istituto Italiano di Tecnologia (IIT), Center for Biomolecular Nanotechnologies, Via Barsanti, 73010 Arnesano, Italy.}
\affiliation{Istituto Nanoscienze-CNR, Euromediterranean Center for Nanomaterial Modelling and Technology (ECMT), Via per Arnesano 16, 73100 Lecce, Italy.}

\author{Dario Pisignano}
\affiliation{Department of Mathematics and Physics ``Ennio De Giorgi'', University of Salento, Via per Arnesano, 73100, Italy.}
\affiliation{Istituto Nanoscienze-CNR, Euromediterranean Center for Nanomaterial Modelling and Technology (ECMT), Via per Arnesano 16, 73100 Lecce, Italy.}

\author{Cristian Cirac\`i}
\affiliation{Istituto Italiano di Tecnologia (IIT), Center for Biomolecular Nanotechnologies, Via Barsanti, 73010 Arnesano, Italy.}
\email[]{cristian.ciraci@iit.it}

\begin{abstract}
Efficient manipulation and long distance transport of single-photons is a key component in nanoscale quantum optics.
In this letter, we study the emission properties of an individual light emitter placed into a nanofiber and coupled to a metallic nanoparticle. 
We find that plasmonic field enhancement together with the nanofiber optical confinement uniquely and synergistically contribute to an overall increase of  emission rates as well as quantum yields.
\end{abstract}


\maketitle

Being able to transport single-photons over long distances is a primary requirement towards integrated quantum technology. 
A nanoscale control of the emission, coupling and transport of single-photons could greatly extend the reach of quantum cryptography~\cite{beveratos_single_2002} 
and enable quantum information~\cite{bennett_quantum_2000} and computing applications~\cite{knill_scheme_2001}.
It has been shown recently that suitably engineered plasmonic systems can drastically increase decay rates as well as emission directionality 
and far-field coupling~\cite{akselrod_leveraging_2015, akselrod_probing_2014}.

The sub-wavelength nature of plasmonic modes allows, in fact, an efficient coupling between quantum emitters and photonic modes, although strong ohmic losses in plasmonic waveguides 
limit their use in most of practical scenarios.
On the other hand, dielectric waveguides allow lossless propagation but they suffer from coupling efficiency. 
Near-field coupling is fundamentally upper bounded to about a 30\% in silica nanofibers \cite{Klimov:2004is,Fujiwara:2011gs,Yalla:2012gv}.
An efficient configuration would require the emitter to be placed inside the fiber.
Although this is a quite challenging approach, Gaio and co-authors~\cite{gaio_modal_2015} have recently reported a broadband single-photon generation and transport from 
isolated quantum dots embedded in the core of free-standing and sub-wavelength polymer nanofibers. These nanostructures are made of a polymer matrix which is processed from solution in order to realize elongated filaments. Depending on the production method, the resulting system can be an organic or hybrid organic-inorganic nanorod with length in the range of a few to tens of micrometers, or a continuous spun nanofiber \cite{Xia:2003te,Kim:2011ke,Persano:2015jg}. Technologies used in this field include self-assembly assisted by polymer deposition through poor solvents, polymerization methods, nanofluidics, and, notably, electrospinning which is the most promising in terms of amount of produced nanomaterials, low cost, and operational convenience \cite{Reneker:1996uy}. The electrospinning is based on the application of an intense electric field to a solution with sufficient amount of polymer entanglements, and allows for realizing continuous nanofibers whose transversal size well-matches near-UV, visible, or near-IR wavelengths. Furthermore, nanoparticles of virtually any composition \cite{Zhang:2014bs} as well as light-emitting dopants can be added to the spun solutions thus being transferred to the resulting nanofibers and leading to hybrid photonic systems.    

\begin{figure}[hbt]
    \begin{center}
         \includegraphics[width=0.38\textwidth]{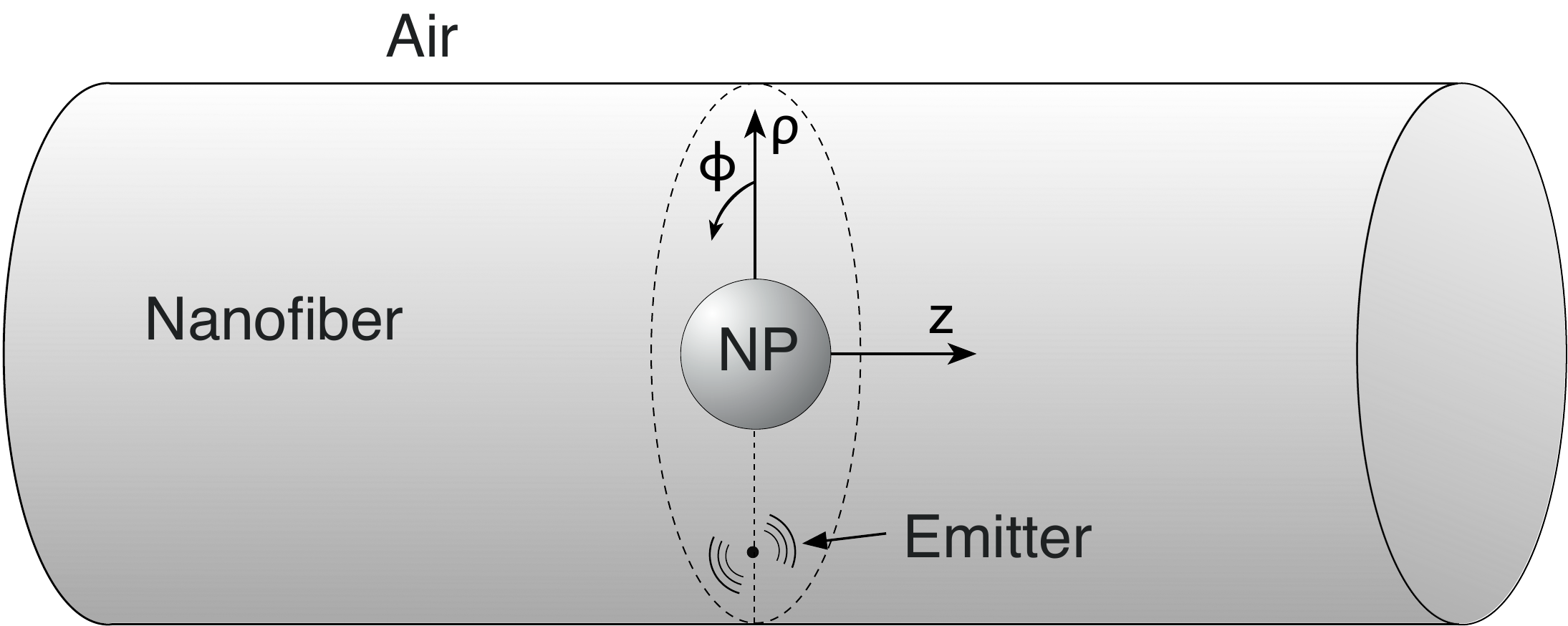}
    \end{center}
    \caption{Illustration of the nanoparticle embedded in a nanofiber system: a dielectric cylinder with a metal sphere at its center, and a dipole located in the central transverse plane. The dipole's position is restricted to the dashed line. Directions are defined by the axes.}
    \label{fig:fiber}
\end{figure}

Metallic inclusions inside these polymeric fibers are especially interesting in this respect. For instance, Au nanorods in poly(vinyl alcohol) nanofibers have been proposed as building blocks of flexible substrates for surface-enhanced Raman scattering \cite{Zhang:2011gq}. Metallic nanoparticles in nanofibers also bring a possibility to directly enhance the emission rates of chromophores inside the optical transport channel. In this letter, we numerically investigate the emission properties of a single emitter placed into a nanofiber and coupled to a metallic nanoparticle (NP). Previous theoretical investigations considered only a pure dielectric nanofiber (i.e. without the NP), see e.g. Ref. \cite{zako00}.

We analyze the emission properties as a function of the distance from the NP and show that the decay rates are increased with respect to the case where NPs are absent.
We compare our results to the dipole emission nearby a NP in an unbounded medium and in the nanofiber in the absence of NP inclusions, and 
show that the combination of NP and polymeric fiber uniquely enhances the dipole emission.
Although here we present a purely numerical analysis, our results may be of great interest for ongoing and future experiments.

In order to evaluate simultaneously the effect of the fiber and NP plasmonic behavior we perform full-wave simulations of a section of the nanofiber modeled as a 
dielectric cylinder of permittivity $\varepsilon$. 
Polymer nanofibers might display an anisotropic permittivity reaching values as high as $\varepsilon=9$~\cite{gaio_modal_2015}, however, here we will consider the isotropic case.
A metal sphere is placed at the fiber center, and a dipole emitter is located on the fiber transverse cross-section plane in correspondence to the center of the NP as shown in Fig.~\ref{fig:fiber}.
We are interested in the modification of the luminescence properties of the emitter as a result of the confinement inside the nanofiber and in proximity of the metal sphere.

We assume that the emitter is weakly coupled to its environment allowing us to model it as a two-level system with transition dipole moment~$\mathbf{p}$ and transition frequency~$\omega$. 
The luminescence properties depend on its position~$\mathbf{r}_{\text{m}}$ and its orientation denoted by the dipole moment unit vector~$\mathbf{\hat{n}}_p$, and they can be calculated by 
evaluating the spontaneous emission rate~$\gamma_{\text{sp}}= \gamma_{\text{r}} + \gamma_{\text{nr}} + \gamma_{\text{int}}^0 $~\cite{ciraci_numerical_2014, novotny_principles_2012},
where~$\gamma_{\text{r}}$ represents the radiative emission rate accounting for emitted photons, $\gamma_{\text{nr}}$~is the non-radiative emission rate 
accounting for the energy dissipated in the environment and~$\gamma_{\text{int}}^0$, representing the emitter's internal 
non-radiative emission rate accounting for intrinsic decay such as phononic or trapped modes, which does not depend on the electromagnetic environment.
Fermi's golden rule lets us compute the total decay rate~$\gamma_{\text{sp}}$ as:
\begin{equation}
\gamma_{\text{sp}} = \frac{2 \omega^2}{\hbar \varepsilon_0  c^2} \left| \mathbf{p} \right| 
\left[ \mathbf{\hat{n}}_p \cdot \mathrm{Im} \left\{ \mathbf{G} ( \mathbf{r}_{\text{m}}, \mathbf{r}_{\text{m}} ) \right\} \cdot \mathbf{\hat{n}}_p \right]  + \gamma_{\text{int}}^0,
\end{equation}
where~$\mathbf{G}$ is the Green dyadic function of the system evaluated at the location of the dipole.
In lossy systems, such as plasmonic NPs, the contribution of unwanted non-radiative states to the total emission can be important, and it generally produces strong enhancements of the overall $\gamma_{\text{sp}}$. 
A more relevant parameter quantifying the probability that a photon is actually emitted (and not re-absorbed) 
during a dipole transition is the quantum yield $q ( \mathbf{r}_{\text{m}}, \mathbf{\hat{n}}_p, \omega ) = {\gamma_{\text{r}}}/{\gamma_{\text{sp}}}$. The radiative emission rate $\gamma_{\text{r}}$ can be
 obtained by difference as $\gamma_{\text{r}}=\gamma_{\text{sp}}-\gamma_{\text{nr}}-\gamma_{\text{int}}^0$, where the 
absorption occurring in the domain~$\Omega$ containing the system gives us directly $\gamma_{\text{nr}}$ as:
\begin{equation}
    \gamma_{\text{nr}} = \frac{1}{2} \frac{\gamma_{\text{r}}^0}{W_{\text{r}}^0} \int_{\Omega} \mathrm{Re} \left\{ \mathbf{J}^* \cdot \mathbf{E} \right\} \mathrm{d} V \, ,
\end{equation}
where~$\gamma_\text{r}^0={\omega^3 \left| \mathbf{p} \right|^2}/\left({3 \hbar \pi \varepsilon_0 c^3}\right)$ is the radiative emission rate 
in free space and~$W_{\text{r}}^0={\omega^4 \left| \mathbf{p} \right|^2}/\left({12 \pi \varepsilon_0 c^3}\right)$ the total power radiated in free space.
Finally, the value of the internal emission rate can be obtained from the emitter's quantum yield in free space,~$q^0$, as $\gamma_{\text{int}}^0 = \gamma_{\text{r}}^0 \left( {1}/{q^0} - 1 \right)$.

The electromagnetic properties of the whole system, constituted by the nanofiber, the NP and the dipole, were calculated  using a commercially available software
 based on the finite-element method, namely \textsc{Comsol} Multiphysics~\cite{comsol}.
The three dimensional geometry sketched in Fig.~\ref{fig:fiber} features symmetries that reduce the simulation volume to one fourth of the system's total volume.
Two planes of symmetry going through the center of the fiber were used: a plane transverse with respect to the fiber, and a longitudinal plane perpendicular to the first one.
The dipole is located on the intersection of the two planes of symmetry and can have different orientations, each requiring different boundary conditions.
A boundary plane on which the dipole is located is considered as a perfect electric conductor or perfect magnetic conductor if the dipole is perpendicular to the plane or along the plane, respectively.
A 150 nm wide layer of air is added around the fiber in the radial direction to take into account reflections of the 
fields at the interface, but not at the extremities of the cylinder to avoid Fabry-Perot reflections, which is to say that we therefore assume having an infinitely long fiber.
The outer boundary of the domain is implemented with perfectly matched layers \cite{Berenger:1994eu} that prevent reflections back inside the domain. The quantum emitter is modeled as a dipolar current whose amplitude is chosen such that its radiated power in vacuum is given by $W_{\text{r}}^0$ for the same
 dipole moment~${\mathbf{p}}$.

\begin{figure}
    \begin{center}
        \includegraphics[width=0.47\textwidth]{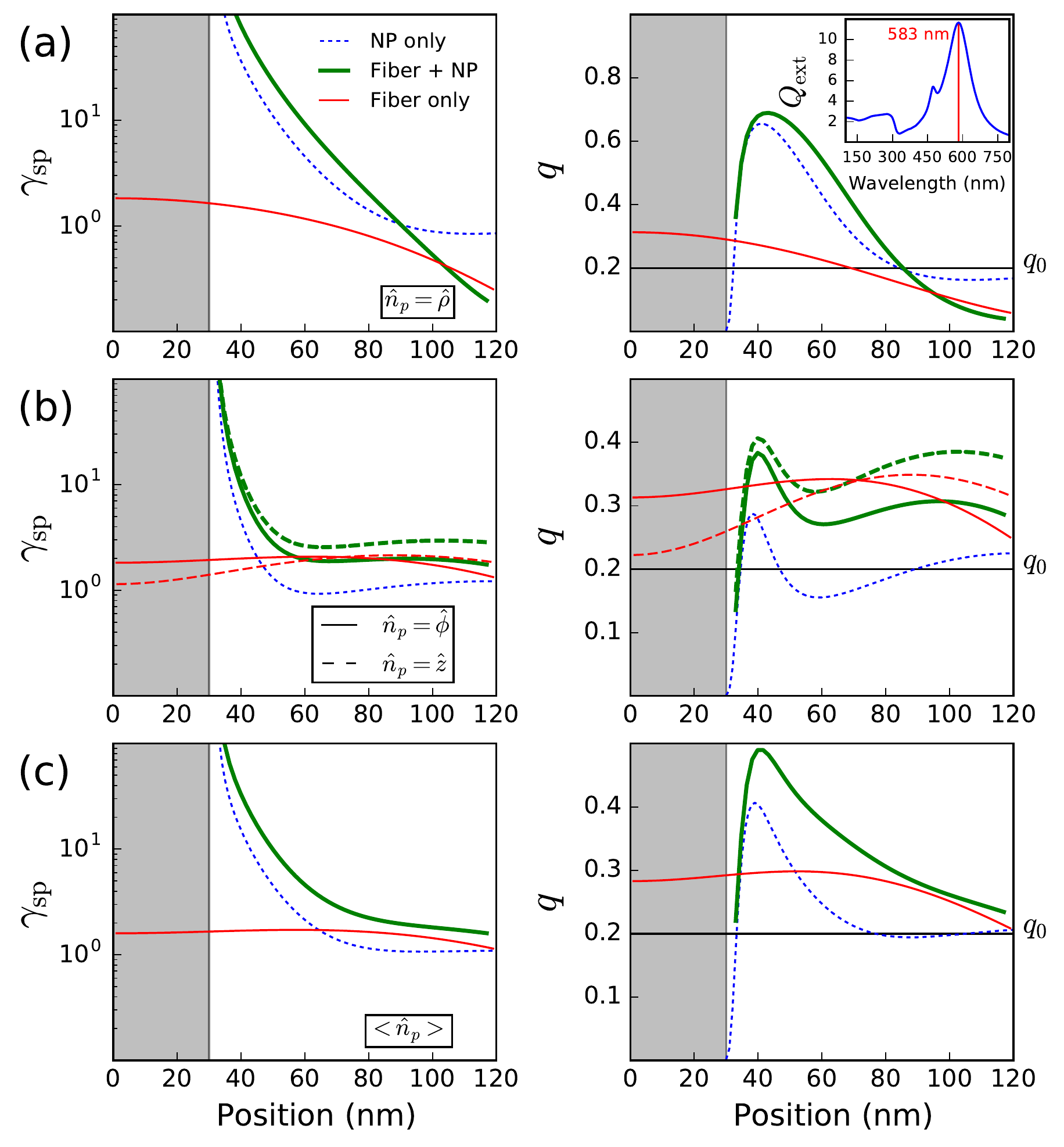}
    \end{center}
    \caption{Results for $R_{\text{fiber}} = 120$ nm.
    (a)~Spontaneous emission rate (left column) and quantum yield (right column) at the LSPR wavelength for a dipole along~$\hat \rho$; see text for details.
    Inset shows the extinction efficiency, the LSPR wavelength is~$\lambda = 583$ nm with~$\varepsilon = 4$.
    (b)~Same as in (a) but for a dipole along~$\hat\phi$ or~$\hat z$. 
    (c)~Same as in (a) but the dipole's orientation is averaged.}
    \label{fig:r120nm}
\end{figure}

In order to understand the impact of the fiber and of the NP on the emitter properties we  have  carried out calculations for different fiber permittivities, different fiber radii and with a varying distance between the dipole and the metal sphere.
We considerd a silver NP with a permittivity given by Palik~\cite{palik_handbook} and with a radius of~30 nm . The nanofiber permittivity is set to $\varepsilon = 4$ and the dipole emission frequency to the NP's extinction spectrum peak, i.e. the localized surface plasmon resonance~(LSPR) at $\lambda = 583 nm$ as shown in the inset of Fig.~\ref{fig:r120nm}a.
The quantum efficiency of the emitter in free space is~$q_0 = 0.2$, which is the value of experimentally commonly used fluorophores (e.g. Cy5)\cite{Rose:2014ko}.

We consider first a fiber of radius $R=120 nm $ such that the single-mode condition\cite{Saleh2007} $V=k_0 R\sqrt{\varepsilon-1} < 2.405$ is satisfied, and model a section of the fiber of length ~3 $\mu m$ with the NP at its center. In Fig.~\ref{fig:r120nm}a we show the resulting spontaneous emission rate~$\gamma_\text{sp}$ (left column) and the quantum yield~$q$ (right column) for a dipole 
perpendicularly polarized with respect to the NP surface as a function of its distance from the NP; see the ``Fiber+NP'' thick-green curves.

The simulation results are compared to the exact solution of a dipole emitting near a metallic sphere in an unbounded medium\cite{anger_enhancement_2006,kim_classical_1988, mertens_plasmon-enhanced_2007}; see the ``NP only'' blue-dashed curves.
Simulations of a nanofiber with a dipole but with no NP have also been carried out (see the ``Fiber only'' thin-red curves) to quantify the role of the nanofiber dielectric confinement.

We find that confining the NP and the dipole in a fiber clearly leads to several interesting effects:

\begin{figure}
    \begin{center}
        \includegraphics[width=0.47\textwidth]{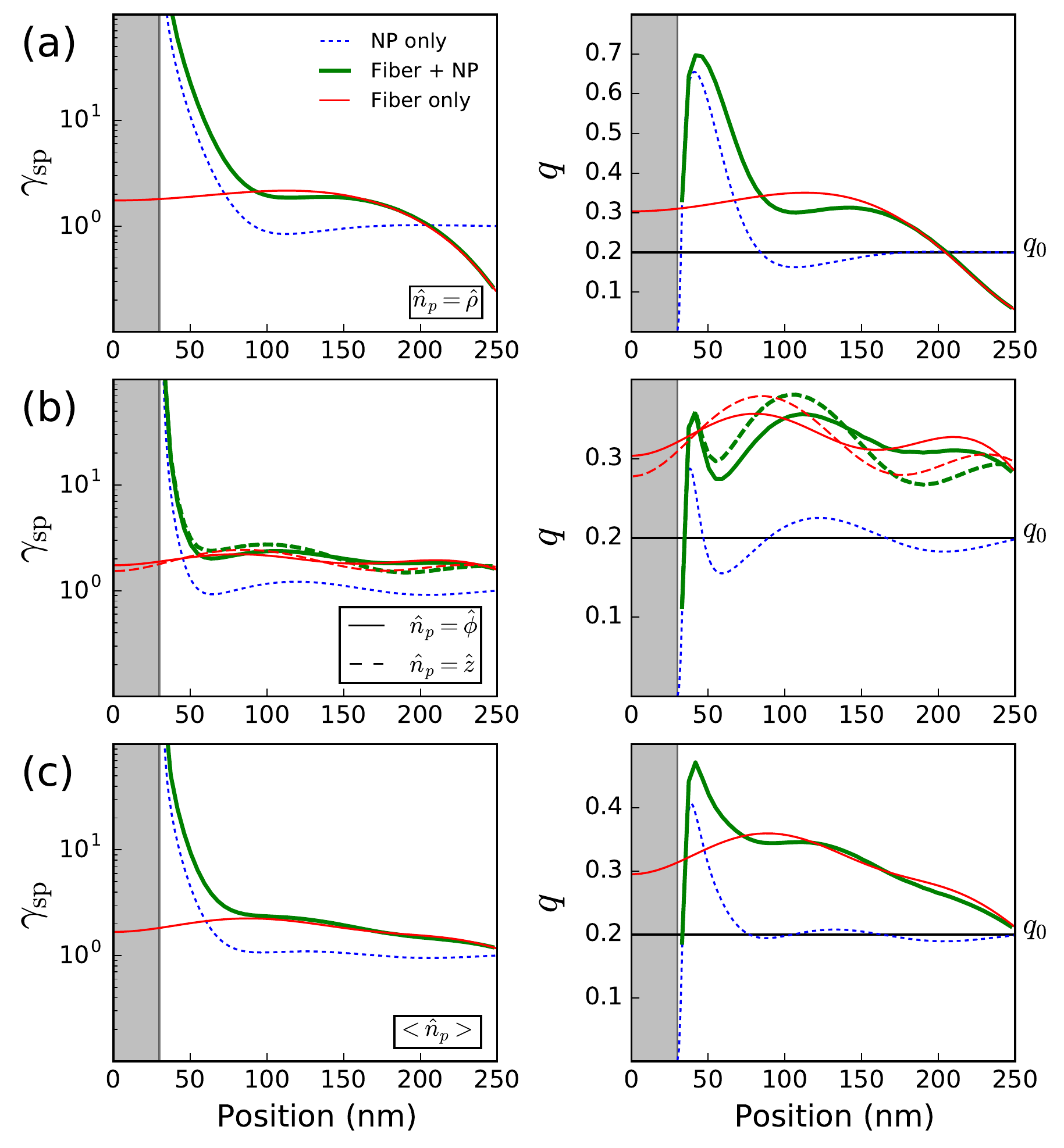}
    \end{center}
    \caption{Results for $R_{\text{fiber}} = 250 nm$.
    (a)~Spontaneous emission rate (left column) and quantum yield (right column) at the LSPR wavelength for a dipole along~$\hat \rho$.
    (b)~Same as in (a) but for a dipole along~$\hat\phi$ or~$\hat z$. 
    (c)~Same as in (a) but the dipole's orientation is averaged.}
    \label{fig:resonance}
\end{figure}

(i) For  distances up to about~70 nm the ``Fiber+NP'' spontaneous emission rate as well as the quantum yield have a similar profile as the ``NP only'' system, but the former is larger. For $d \approx 40 nm$ the maxium quantum yield is up to~$3.5$ ($3.3$) times larger than in free space for the ``Fiber+NP'' (``NP only'') system. Note that for very short distances, $d \approx 30 nm$, the usual quenching is found in both cases;

(ii) As the dipole gets close to the fiber's edge, we can observe a decrease of the spontaneous emission rate that undergoes a reduction due to both the NP and the fiber confinement, as it can be seen by the red curves in Fig.~\ref{fig:r120nm}a;

(iii) If the dipole is parallel with respect to the NP surface, hence it is oriented along~$\hat\phi$ or~$\hat z$, there is a similar increase (going from  ``NP only''  to ``Fiber+NP'') in the spontaneous emission rate and quantum yield, see Fig.~\ref{fig:r120nm}b.  However, in this case, the maximum quantum yield is only up to~$2.0$ times larger than in free-space. Moreover, the decrease near the edge of the fiber that occurs for a dipole oriented along~$\hat\rho$ is not observed.

Finally, in Fig.~\ref{fig:r120nm}c we report the results when the three dipole orientations have been averaged. 
This case reflects more closely the behavior expected in experimental conditions where there is no control on the orientation of the dipole of a chromophore embedded in the polymer filament.
The effect of the NP in the fiber can be clearly observed up distances of about~80 nm. 
For the quantum yield this increase is up to~$2.5$ times larger than in free-space, and it is up to $1.6$ ($1.2$) larger
than in ``Fiber only'' (``NP only'') case. 
The quenching observed at the edge of the fiber for a dipole along~$\hat\rho$ 
is compensated for by the other orientations for which it does not occur.

\begin{figure*}[hbt]
    \begin{center}
         \includegraphics[width=0.95\textwidth]{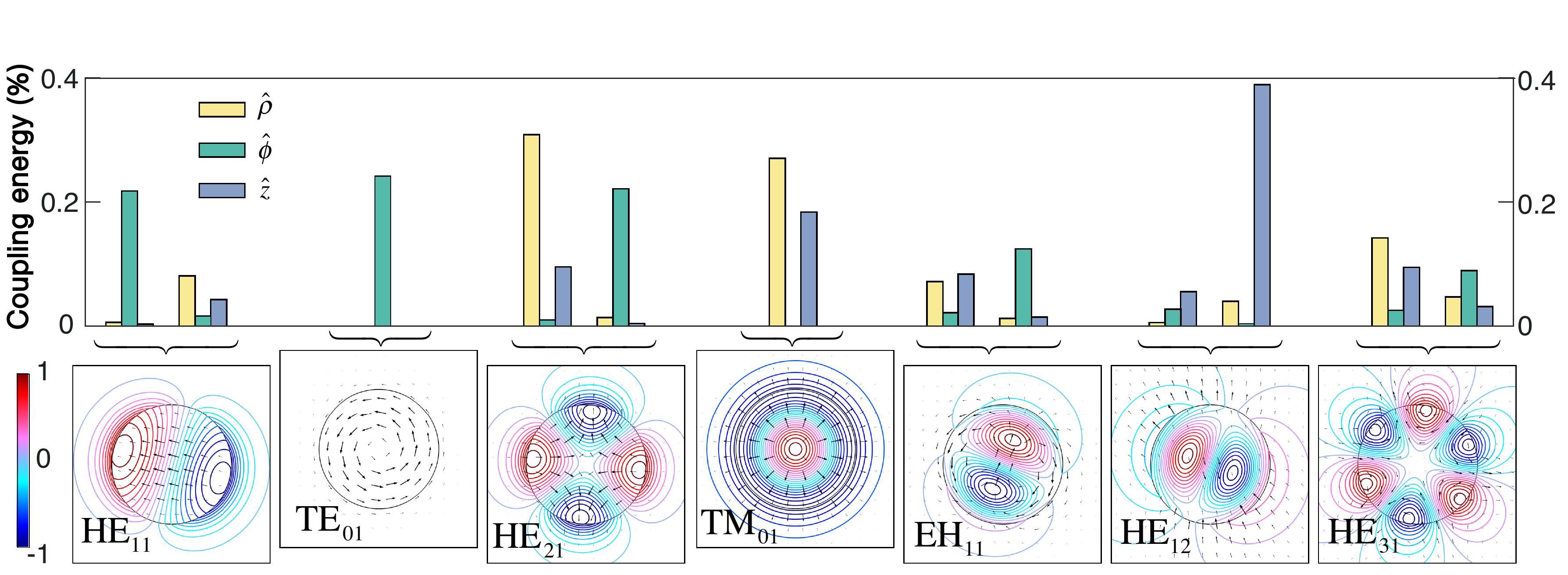}
    \end{center}
    \caption{Modes supported by the waveguide and corresponding coupling energies 
for an emitting dipole oriented along ${\hat n}_p={\hat\rho}, {\hat\phi}, {\hat z}$ and positioned at $45$ nm from an Ag nanoparticle. Arrows represent transverse field components, 
while the contour plot depicts the $z$-component. HE and EH modes are degenerate with degree equal to 2. Degenerate modes are not reported.}
    \label{fig:modes}
\end{figure*}

Since using single-mode fibers in experimental conditions is not always possible or convenient, we analyze also the case of a multi-mode fiber with a larger radius~$ R =250 nm$.
The results are reported in Fig.~\ref{fig:resonance}. A similar overall behavior occurs, and very comparable maximal values of quantum yield enhancement are reached.
We also report  in Fig.~\ref{fig:modes}  all the modes supported by the fiber and their respective coupling energies.
The modes were calculated using the eigenvalue solver of \textsc{Comsol} Multiphysics\cite{comsol} and refer to the bare fiber with no NP inclusions.
The coupling energies, reported as a bar plot in Fig.~\ref{fig:modes}, refer to a dipole placed at $d=R/2$ from the center of the fiber in the presence of a NP, and oriented along $\hat\rho$, $\hat\phi$ and $\hat z$ respectively.
These coefficients, $|a_n|^2$, can be calculated using the orthogonality of the modes as\cite{jackson}:
\begin{equation}
{a_n} = \iint {\left( {{\bf{E}} \times {\bf{H}}_n^*} \right) \cdot {\bf{\hat z}}} \;dx dy, 
\end{equation}
where ${\bf H}_n$ is the magnetic field of the $n$-th mode and ${\bf E}$ is the electric field emitted by the dipole inside the fiber taken at the cross-section plane 7 $\mu$m away form the emitter.
The coefficients are normalized so that $\sum_{n} |a_n|^2=1$, for each dipole orientation.

For a dipole oriented along~$ \hat{r} $, it can be seen from Fig.~\ref{fig:resonance}a that although the behavior is otherwise very similar, there is a plateau between the strong enhancement near the NP and the quenching near the fiber's edge. The HE$_{21}$ and TM$_{01}$ modes are mainly excited and both have weak radial components near the edge. The quantum yield is up to $3.5$ times higher than in vacuum.


When the dipole is oriented along~$ \hat{\phi} $ or~$ \hat{z} $, the curve is shaped very similarly to the ``NP only'' case but simply shifted upwards due to the confinement in the fiber. The quantum yield is up to $1.9$ times higher than in vacuum.

When the dipole orientation is averaged, it is clear that the plasmonic effects are only noticeable at very close distances from the NP, up to~$ d \approx 70 nm $, where they add to the enhancement due to the confinement alone. At larger distances, the curves for ``Fiber-only'' and ``Fiber+NP'' overlap and the confinement in the fiber is responsible for the enhancement. For the quantum yield this enhancement is up to~$2.4$ times larger than in free-space.

Lastly, simulations have also been carried out for a fiber with~$\varepsilon=2.5$, $ R = 250 nm$ and with a dipole emitting at the LSPR which is then~484 nm (data not reported). The profile of the curves is similar to Fig.~\ref{fig:resonance}, but the enhancement is lower overall.
The quantum yield enhancement is up to~$3.0$, $1.5$ and~$1.9$ times larger than in free-space for a dipole with a polarization 
that is respectively perpendicular, parallel and averaged with respect to the NP's surface.

In conclusion, we have described theoretically the emission properties of a dipole inside a fiber with a NP at its center, with the emitter 
located in the same transverse plane as the metal sphere.
We observed that the confinement inside the fiber alone increases the spontaneous emission rate and quantum yield compared to a metal sphere in an unbounded 
medium, unless the dipole is oriented in the radial direction and close to the fiber's edge in which case the radiative emission rate of the dipole is reduced.
In the close vicinity of the metal sphere, the usual strong plasmon-enhanced luminescence and quenching are observed, with the enhancement being stronger 
than in an unbounded medium.

These results could be in principle improved by engineering the plasmonic components, by, for example using metallic dimers or chains, instead of single particles. 
However this approach is still quite challenging to realize experimentally. 
We believe that our analysis shows enhancement of photon-emission rates that can be practically achieved and will constitute a guide for future experiments.

\begin{acknowledgements}
The research leading to these results has received funding from the
European Research Council under the European Union's Seventh Framework
Programme (FP/2007-2013)/ERC Grant Agreement n. 306357 (``NANO-JETS'').
\end{acknowledgements}

\end{document}